\documentclass[manuscript,screen]{acmart}

\AtBeginDocument{%
  }

\setcopyright{acmlicensed}
\copyrightyear{2024}
\acmYear{2024}
\acmDOI{XXXXXXX.XXXXXXX}

\acmISBN{978-1-4503-XXXX-X/18/06}

\usepackage{graphicx}
\usepackage{multirow}
\usepackage{booktabs}
\usepackage{subcaption}
\usepackage{color}
\usepackage{dblfloatfix}
\usepackage{tikz-qtree}
\usepackage{listings}
\usepackage{draftwatermark}

\SetWatermarkText{Preprint Under Review}
\SetWatermarkScale{0.3}
\begin{document}

%%
%% The "title" command has an optional parameter,
%% allowing the author to define a "short title" to be used in page headers.
\title{Context-aware Code Summary Generation}

%%
%% The "author" command and its associated commands are used to define
%% the authors and their affiliations.
%% Of note is the shared affiliation of the first two authors, and the
%% "authornote" and "authornotemark" commands
%% used to denote shared contribution to the research.
\author{Chia-Yi Su}
%\authornote{Both authors contributed equally to this research.}
\email{csu3@nd.edu}
\affiliation{%
  \institution{University of Notre Dame}
  \city{Notre Dame}
  \state{Indiana}
  \country{USA}
}

\author{Aakash Bansal}
\email{abansal@lsu.edu}
\affiliation{%
  \institution{Louisiana State University}
  \state{Baton Rouge}
  \country{USA}}

\author{Yu Huang}
\email{yu.huang@vanderbilt.edu}
\affiliation{%
  \institution{Vanderbilt University}
  \state{Nashville}
  \country{USA}
}

\author{Toby Jia-Jun Li}
\email{toby.j.li@nd.edu}
\affiliation{%
 \institution{University of Notre Dame}
 \state{Notre Dame}
 \country{USA}}

\author{Collin McMillan}
\email{cmc@nd.edu}
\affiliation{%
  \institution{University of Notre Dame}
  \state{Notre Dame}
  \country{USA}}

%%
%% By default, the full list of authors will be used in the page
%% headers. Often, this list is too long, and will overlap
%% other information printed in the page headers. This command allows
%% the author to define a more concise list
%% of authors' names for this purpose.
\renewcommand{\shortauthors}{Su et al.}

%%
%% The abstract is a short summary of the work to be presented in the
%% article.
\begin{abstract}
  Code summary generation is the task of writing natural language descriptions of a section of source code.  Recent advances in Large Language Models (LLMs) and other AI-based technologies have helped make automatic code summarization a reality.  However, the summaries these approaches write tend to focus on a narrow area of code.  The results are summaries that explain what that function does internally, but lack a description of why the function exists or its purpose in the broader context of the program.  In this paper, we present an approach for including this context in recent LLM-based code summarization.  The input to our approach is a Java method and that project in which that method exists.  The output is a succinct English description of why the method exists in the project.  The core of our approach is a 350m parameter language model we train, which can be run locally to ensure privacy.  We train the model in two steps.  First we distill knowledge about code summarization from a large model, then we fine-tune the model using data from a study of human programmer who were asked to write code summaries.  We find that our approach outperforms GPT-4 on this task.
\end{abstract}

%%
%% The code below is generated by the tool at http://dl.acm.org/ccs.cfm.
%% Please copy and paste the code instead of the example below.
%%


%%
%% Keywords. The author(s) should pick words that accurately describe
%% the work being presented. Separate the keywords with commas.

\received{20 February 2007}
\received[revised]{12 March 2009}
\received[accepted]{5 June 2009}

%%
%% This command processes the author and affiliation and title
%% information and builds the first part of the formatted document.
\maketitle
\section{Introduction}

A code summary is a short description in natural language of a section of source code~\cite{haiduc2010supporting}.  Automated generation of code summaries has long been considered a ``holy grail'' of software engineering research in some circles~\cite{allamanis2018survey}, due to the intense market need and yet technical challenge of machine understanding of code.  The market need is due to the high value programmers place on good documentation combined with a notorious lack of time to write that documentation~\cite{robillard2017demand}.  The technical challenge is due to the complexity of building machine representations of semantic meaning of code combined with writing natural language itself.  The technical challenge has been described for decades as the ``concept assignment problem''~\cite{biggerstaff1993concept} and recent advancements in AI-based technology such as Large Language Models (LLMs) have helped make practical solutions a reality.

The vast majority of code summarization solutions described in the literature focus on comments for well-defined sections of code, such as Java methods~\cite{chen2021why}.  This focus is encouraged by platforms such as JavaDoc~\cite{kramer1999api} and documentation style guides from large corporations~\cite{rani2021what}.  Existing approaches tend to generate summaries that describe the internals of these sections of code, such as the algorithms they implement or the inputs and outputs of a method.  These approaches have, in recent years, almost all been based on neural models trained with example code summaries~\cite{zhang2022survey}.  Progress has been made from novel neural architectures combined with ever-larger model size and data sizes.

A gap in the ability of current approaches is in describing \emph{why} code exists in a program.  ``Why'' information refers to the purpose of the program and the role that specific code plays in that purpose~\cite{roehm2012professional, ko2007information}. This information can provide the background knowledge to help programmers understand the big picture of functions~\cite{chen2021why}. 
Recent advancements show the accumulated progress of years of research: it is now possible to give an LLM a Java method and a prompt asking for a summary, and receive output comparable to summaries written by humans~\cite{su2024distilled}.  However, as we will show in this paper, these models struggle to describe the purpose behind the method in the context of a whole program.  %This gap in ability may reflect a lack of task-specific training as LLMs are capable of other coding tasks~\cite{nam2024using}.% and now often have large input lengths (e.g., 128k tokens in GPT-4).%  The ingredients for ``context aware'' code summary generation exist, but have not been fully formed together.

In this paper, we present a novel language model-based approach for context-aware code summary generation of Java methods.  By ``context aware'', we mean summaries that explain why a method exists in a program.  Given a target method for which a summary is needed, our approach works by: 1) finding the methods in the program that call the target method, 2) generating a summary of what each caller method does, using existing code summarization techniques, 3) generating a new summary of the target method using the summaries of the caller methods as a guide.  We implement this idea first using GPT-4, to demonstrate its feasibility.  Then we re-implement the idea on a smaller model with 350m parameters.  The small model is helpful because data privacy rules can insist that proprietary code not be sent to third-parties, which LLMs such as GPT-4 require while small models can be run locally~\cite{derner2023beyond}.%  Second, we can fine-tune our model so that it better conforms to examples of exemplary context-aware summaries provided by human experts.

An advantage to our small model is that we can fine-tune the model so that it better conforms to examples of exemplary context-aware summaries provided by human experts.  We collect 256 examples of summaries written by human experts from an experiment in which the people were asked to thoroughly read a Java project and write summaries for individual methods.  We adapt a fine-tuning approach in which we froze select parameters of the model to reduce overfitting leading to model hallucinations.

We evaluate our approach in a qualitative study involving six experiments.  In one set of three experiments, we study our approach versus baselines using the commercial LLMs Gemini and GPT-4.  In a second set of three experiments, we study our approach using the small language model that we fine-tune ourselves, and compare that small model to the best-performing commercial LLM as well as human-written references.  Sixty professional programmers participated in our experiments.  We show that our small model outperforms both Gemini and GPT-4 on the specific task of context-aware code summary generation.

%We evaluate our approach in a quantitative study using automated metrics and a qualitative study with human programmers.  In the quantitative study, we compare the similarity of the generated summaries to the summaries written by human experts with metrics such as METEOR and USE.  In the qualitative study, we solicit feedback from human programmers (different people than those who wrote the summaries we used for fine-tuning).  We show across various measures that our model outperforms GPT-4 on the specific task of context-aware code summary generation. %YH: you mean METEOR and BLEU?

%YH: Intro is very well written! One thing I would like to point out: you advertised early that previous work does not consider "why" code exists in a program - which is great! However, I cannot see whether your approach can actually explain "why" - can the METEROR and USE and/or the human evaluation indicate your summaries are indeed better on explaining "why this code snippet exist in the program"than GPT4? "outperforms GPT4" is a bit vague. How did you measure that? If you did measure "why", then here you want to highlight it (don't sell yourself short :)). If you didn't measure "why", then the intro could be a bit risky cause you introduced a challenge that previous work failed to solve, but your approach cannot solve it either...
\section{Background and Related Work}
\label{sec:bg}

\begin{table}[!b]
\vspace{-7mm}
	%\begin{table}[t!]
	{\small
		%\vspace{-0.5cm}
		\begin{tabular}{p{4.9cm}p{0.6cm}p{0.6cm}p{0.6cm}}
			 & G          & L & C                \\
			%\textcolor{white}{*}McBurney~\emph{et al.}~(2014)~\cite{mcburney2014automatic}						& x &  	&   &   & x \\
			%\textcolor{white}{*}Zhang~\emph{et al.}~(2016)~\cite{zhang2016towards}					& x &  	&   &   & x \\
			%\textcolor{white}{*}Iyer~\emph{et al.}~(2016)~\cite{iyer2016summarizing}				&   & x &   &   &   \\
			%\textcolor{white}{*}Rodeghero~\emph{et al.}~(2017)~\cite{rodeghero2017detecting}		& x &  	&   &   & x \\
			%\textcolor{white}{*}Fowkes~\emph{et al.}~(2017)~\cite{fowkes2017autofolding}			& x &  	&   &   &   \\
			%\textcolor{white}{*}Badihi~\emph{et al.}~(2017)~\cite{badihi2017crowdsummarizer}		& x &  	&   &   &   \\
			%\textcolor{white}{*}Loyola~\emph{et al.}~(2017)~\cite{loyola2017neural}				&   & x	&   &   &   \\
			%\textcolor{white}{*}Lu~\emph{et al.}~(2017)~\cite{lu2017learning}						&   & x	&   &   &   \\
			%\textcolor{white}{*}Jiang~\emph{et al.}~(2017)~\cite{jiang2017automatically}			&   & x	&   &   &   \\
			%\textcolor{white}{*}Hu~\emph{et al.}~(2018)~\cite{hu2018summarizing}					&   & x	&   &   &   \\
			%\textcolor{white}{*}Hu~\emph{et al.}~(2018)~\cite{hu2018deep}							&   & x	& x &   &   \\
			%\textcolor{white}{*}Allamanis~\emph{et al.}~(2018)~\cite{allamanis2018learning}		&   & x	& x &   &   \\
			%\textcolor{white}{*}Wan~\emph{et al.}~(2018)~\cite{wan2018improving}					&   & x	&   &   &   \\
			%\textcolor{white}{*}Liang~\emph{et al.}~(2018)~\cite{liang2018automatic}				&   & x	&   &   &   \\
			\textcolor{white}{*}Alon~\emph{et al.}~(2019)~\cite{alon2019code2seq, alon2019code2vec}	& x &   &   \\
			%\textcolor{white}{*}Gao~\emph{et al.}~(2019)~\cite{gao2019neural}						&   &   &   \\
			\textcolor{white}{*}LeClair~\emph{et al.}~(2019)~\cite{leclair2019neural}				& x &   &   \\
			%\textcolor{white}{*}Nie~\emph{et al.}~(2019)~\cite{nie2019framework}					&   &   &   \\
            \textcolor{white}{*}Wei~\emph{et al.}~(2019)~\cite{wei2019code}					        &   &   &   \\
			\textcolor{white}{*}Haldar~\emph{et al.}~(2020)~\cite{haldar2020multi}					&   &   &   \\
			\textcolor{white}{*}Ahmad~\emph{et al.}~(2020)~\cite{ahmad2020transformer}				&   &   &   \\
			\textcolor{white}{*}Haque~\emph{et al.}~(2020)~\cite{haque2020improved}					&   &   & x \\ 
			\textcolor{white}{*}Z{\"u}gner~\emph{et al.}~(2021)~\cite{zugner2021languageagnostic}	& x &   &   \\
			\textcolor{white}{*}Liu~\emph{et al.}~(2021)~\cite{liu2021retrievalaugmented}			& x &   &   \\
			\textcolor{white}{*}Bansal~\emph{et al.}~(2021)~\cite{bansal2021projcon}				&   &   & x \\
            \textcolor{white}{*}Wang~\emph{et al.}~(2021)~\cite{wang2021codet5}				        &   & x &   \\
            \textcolor{white}{*}Choi~\emph{et al.}~(2023)~\cite{choi2023readsum}				    & x &   &   \\
            \textcolor{white}{*}Bansal~\emph{et al.}~(2023)~\cite{bansal2023function}				& x &   & x \\
            \textcolor{white}{*}Gao~\emph{et al.}~(2023)~\cite{gao2023code}				            & x &   &   \\
            \textcolor{white}{*}Zhang~\emph{et al.}~(2024)~\cite{zhang2024eyetrans}				    & x &   &   \\
            \textcolor{white}{*}Ahmed~\emph{et al.}~(2024)~\cite{ahmed2024automatic}				&   & x &   \\
            \textcolor{white}{*}Pan~\emph{et al.}~(2024)~\cite{pan2024mesia}				        &   &   & x \\
            \textcolor{white}{*}Ding~\emph{et al.}~(2024)~\cite{dingcode}				            &   & x &   \\
            \textcolor{white}{*}Su~\emph{et al.}~(2024)~\cite{su2024distilled}				        &   & x &   \\
			\textcolor{white}{*}  (This Paper)											            &   & x & x \\
		\end{tabular}
	}
	\vspace{0.1cm}
	\caption{Selection of publications related to this paper from the last five years.  All approaches cited above are based on neural networks in one way or another.  Column $G$ means the code is modeled as a graph.  $L$ means LLM-based designs.  $C$ means learning chiefly from code context.}
	\label{tab:screlated}
	%\vspace{-0.4cm}
	%\end{table}
\end{table}

Source code summarization refers to the task of writing brief natural language ``summaries'' of what code does~\cite{haiduc2010supporting}.  Code summarization is a subset of code comment generation research.  Whereas a code comment may be any type of description for any amount of source code, a ``summary'' intends to focus on the high-level purpose behind a well-defined section of code (e.g., a method or class).  The idea of a summary is to help programmers quickly understand why a section of code exists.  In recent years, approaches to write these summaries have become better at describing the internals of a section of code such as what algorithm is implemented, but the task of describing \emph{why} is still elusive~\cite{chen2021why}.

Table~\ref{tab:screlated} below shows a small selection of the publications most-related to this paper over the last five years.  The heart of almost all recent code summarization approaches is a neural model.  The strategies used in these approaches may be broadly classified into one or more of four groups: 1) improvements to neural model architecture, 2) graph-based representations of code, 3) large language model-based, and 4) code context-based.  Table~\ref{tab:screlated} shows how LLM-based and context-based approaches are appearing, but this paper's niche is in combining them.

\vspace{-0.25mm}
\subsection{Model Architecture and Training Improvements}

Approaches based on improvements to model architecture and training have generally adapting advancements from general-purpose natural language processing (NLP) approaches to the problem of code summarization.  Ahmad~\emph{et al.}~\cite{ahmad2020transformer} helped bring Transformer models from NLP problems to code summarization.  Wei~\emph{et al.}~\cite{wei2019code} presented a technique that improved performance by training a model to do code generation as well as code summarization.  Haldar~\emph{et al.}~\cite{haldar2020multi} describe an approach for learning to map code tokens and natural language tokens to a joint embedding space.  Many more papers in this area exist, as highlighted in various surveys~\cite{song2019survey, zhang2022survey, zhao2020survey}.

%\vspace{-1mm}
\subsection{Graph-based and Other Code Representations}

Another line of improvements has focused on novel representations of source code, which most emphasis placed on graph-based representations.  One frequent strategy is to use an Abstract Syntax Tree (AST) to represent code structure.  Hu~\emph{et al.}~\cite{hu2018deep, hu2018summarizing} demonstrate a special format for ASTs so that the AST can be used as input to a seq2seq model design.  LeClair~\emph{et al.}~\cite{leclair2019neural} present an approach in which separate encoders are used for the AST and code token inputs, and later present a followup based on a graph neural network~\cite{leclair2020improved}.  Gao~\emph{et al.}~\cite{gao2023code} introduce a modified Transformer architecture designed around code structural features rather than only treating code as a sequence of tokens.  Novel representations of code (mostly centered around graph-based representations) are a reliable way to improve code summarization.

\subsection{Language Model Approaches}

The public's gaze at present is largely fixated on LLMs due to advancements such as OpenAI ChatGPT and Github Copilot.  These tools are capable of code summarization when we prompt it appropriately ~\cite{su2024distilled}.  However, different research efforts have explored language model-based code summarization for several years.  The strategy is, essentially, to fine-tune a language model that has been pretrained on a related, but more-generic, task.  Approaches around the CodeT5 family of models are popular~\cite{wang2021codet5}.  CodeT5 and similar models are pretrained on very large repositories of code, and then fine-tuned to perform code summarization.  Controversy surrounds commercial LLMs that require code to be sent to the managing organization via API calls, due to data privacy and other concerns~\cite{hellendoorn2021growing}, leading to efforts to mimic the high performance of commercial LLMs on a smaller scale~\cite{su2024distilled}.  Broadly speaking, it is also related to retrieval-augmented generation in that we find relevant code to aid in summary generation, the idea is relatively unexplored for code summarization~\cite{chen2024benchmarking, li2022survey}.

\begin{figure*}[b!]
    \centering
    \includegraphics[width=0.65\linewidth]{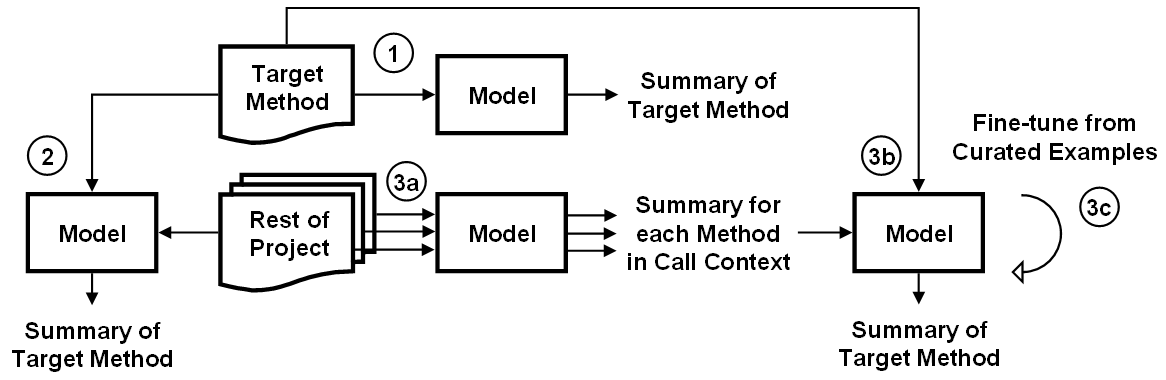}
    \vspace{-2mm}
    \caption{Overview of code summarization processes.  Area 1 is the typical process, with a neural model generating a summary from just the target method itself.  Area 2 is an extended baseline in which a model with a very large input window sees the target method and the rest of the project.  Area 3a-c show our process, which we describe in Section~\ref{sec:approach}.}
    \label{fig:overview}
    \vspace{-2mm}
\end{figure*}

\vspace{-1mm}
\subsection{Code Context}
\label{sec:callcontext}

Code context refers to the other parts of source code surrounding a target area of code.  Code context is sometimes defined as the dependencies and/or dependents of a section of code, such as call relationships~\cite{krinke2006effects}.  The importance of code context for writing summaries has long been recognized and was a fixture of approaches in the era of research prior to neural models~\cite{mcburney2014automatic, song2019survey, sridhara2010towards}.  Different efforts to bring code context into neural models have focused on other methods in the same file~\cite{haque2020improved}, the call graph~\cite{bansal2023function}, and other nearby comments~\cite{pan2024mesia}.  Code context is often considered a crucial component to understanding \emph{why} code exists, since not all the information necessary to write a good summary is always present in the code being summarized itself~\cite{mcburney2014automatic}.  %This paper focuses on bringing the advantages of code context to language model-based approaches.

%YH: again very well written :) One comment: while it is clear the novelty of your approach is to combine LLM and context-aware summarization, it is unclear yet why it is something worth trying. What is specific about LLM that we "must" take advantage of? So important that we cannot miss the oppotunity of combining the two ideas together? 
\section{Approach}
\label{sec:approach}

Our approach is a novel code summarization \textbf{process}, which can be applied using different types of code summarization \textbf{models}. In discussing our approach, we distinguish between code summarization \textbf{processes} and code summarization \textbf{models}. 
A ``process'' refers to the steps required to generate a code summary, including the input/output data and how neural models may be used.  A ``model'' refers to the natural language generation system (usually an autoregressive neural architecture) that reads code or data about code and writes the summary.  %  These days, almost all models are autoregressive neural architectures (see Section~\ref{sec:bg}).
The key novel attributes of our approach are: 1) our approach is designed to describe \emph{why} a method exists using the code context, and 2) our approach supports a relatively small model, which reduces cost and increases data privacy.

\subsection{Baseline Processes}

Figure~\ref{fig:overview} shows an overview of three code summarization processes.  Areas 1 and 2 may be considered baselines called Process 1 and Process 2.  Area 1 shows the process followed by the vast majority of recent code summarization approaches as discussed in Section~\ref{sec:bg}: a neural model that reads a target area of code and generates a summary from only the code in that method.  Area 2 shows the same process, but with the rest of the code from the project also serving as input to the model.  An advantage of this process is that it is possible for producing summaries that include some comprehension of the target code's context.  But a disadvantage is that it requires a model with a very long input window such as GPT-4 which accepts inputs up to 128k tokens.  Later in our experiments, we evaluate our approach (Process 3, described next) against Process 2 instead of Process 1 because Process 1 sees no context and therefore it is not realistic to expect it to describe that context.

%These models can often not be used due to data privacy concerns~\cite{su2024distilled}. 

\subsection{Our Process}

We illustrate our process as three steps (Areas 3a-c) in Figure~\ref{fig:overview}.  First, we use an existing code summarization model to generate a summary of every method in the call context of a target method (Area 3a).  Then we use a second model to generate the summary of the target method, conditioned on the target method itself and the natural language summaries of the methods in the call context (Area 3b).  Finally, we fine-tune the second model using a curated set of gold set examples (Area~3c).  The result is a code summarization system that focuses on answering \emph{why} code exists in the context. In the next two subsections, we show how to use our process using a large commercial and a small open-source model.
%YH: it might need a bit more explanation on the "call context" here. I noticed later in III-C yous aid "we defined call context in II-D" but II-D usues the term "code context" instead of "call context", which can be confusing to reviewers.

\subsection{Our Process via a Commercial Model}
\label{sec:approach_commercial}

We implemented the first two steps (Figure~\ref{fig:overview}, Area 3a-b) of our process in two commercial LLM-based tools: OpenAI's GPT-4~\cite{achiam2023gpt} and Google's Gemini~\cite{team2023gemini}.  Su~\emph{et al.}~\cite{su2024distilled} showed how GPT-3.5 is able to generate code summaries of methods using a simple prompt.  We use the same basic idea to ask the model to produce code summaries for each method in the call context surrounding a target method.  We define call context in Section~\ref{sec:callcontext}.  We use the call context because it has long been shown to be an effective source of information about how code is used~\cite{karrer2011stacksplorer}.  Our prompt form is the following, where {\small\texttt{\{context\}}} is a list of the source code for each method in the call context of the target method:

{\small
\begin{verbatim}
    Write a short description of each of the following
    Java methods, do not duplicate the code in your answer,
    just give a list of the descriptions in paragraph form
    for each description: {context}
\end{verbatim}
}

%YH: I would love to see an example of "context" in your prompt. Maybe considering a motivating example and use it through your paper?  

%The result is a list of short descriptions of each of the methods in the call context.  Next, we send the target method's code (as {\scriptsize\texttt{\{target code\}}}) and the list of descriptions (as {\scriptsize\texttt{\{descriptions\}}}) to the model again, with the prompt:

The result is a list of short descriptions of each of the methods in the call context.  Next, we send the target method's code and the list of descriptions to the model with the prompt:

{\small
\begin{verbatim}
    Consider the following Java method: {target code}
    And consider the following description of Java methods
    that CALL that first Java method: {descriptions}
    Now, write a one-sentence description of WHY the first
    method is used.  The sentence should start with "This
    method is used to".  The WHY description should only 
    include information from the methods that CALL the first
    method and not already in the first method.
\end{verbatim}
}

The result is a one-sentence description of \emph{why} the target Java method exists, which is informed by the descriptions of the methods in the call context.  The call context descriptions give the model high-level information about how the method is used and why it exists in the program.  For example, the purpose for a method that sorts a list of strings is easier to understand if the call context includes methods that show search results to an end-user -- the method sorts strings from a search, so that the user can see them in order.  We show more examples in Section~\ref{sec:examples}.

Note that we do not include the fine-tuning step when using a commercial model.  GPT-4 and Gemini fine-tuning is not available.  OpenAI does provide an API for fine-tuning GPT-3.5, but the details of this process are not public, and we cannot ensure that the data will not be integrated into OpenAI's general model training later.%it is not possible to control how data we upload may be integrated into OpenAI's general model training later.

\subsection{Our Process via an Open-Source Model}
\label{sec:approach_opensource}

We implement all three steps of our process via the open-source model \texttt{jam}~\cite{su2023language} as a foundation.  We used this model because it is pre-trained with Java source code only, has an open-source training set so we can avoid data contamination, and can be run on a single 16GB GPU.  Su~\emph{et al.}~\cite{su2024distilled} have shown the effectiveness of this model for code summarization on par with GPT-3.5 after fine-tuning.  Our implementation required several components:

\subsubsection{Call Context Methods Summary Generation}
\label{sec:approach_callcontext}

This step corresponds to Figure~\ref{fig:overview}, Area 3a.  We used the code summarization model provided by Su~\emph{et al.}~\cite{su2024distilled} which is reported to approximate GPT-3.5 on the code summarization task.  Given a target method, we extract the methods in the call context and send each of those methods to the model using the prompt:
\vspace{-1mm}
{
\begin{verbatim}
    TDAT
    {target method}
    SUMMARY
\end{verbatim}
}
\vspace{-1mm}
This prompt structure is not free-form text like GPT-4 is able to accept.  Instead, it follows the format specified by Su~\emph{et al.}~\cite{su2024distilled} during their fine-tuning process.  The model will output a summary after the SUMMARY header.  Note that the latest model release by Su~\emph{et al.}~\cite{su2024distilled} is a 350m parameter GPT-2 architecture design with a maximum window size of 1024 tokens.  It is not able to accept all call context methods at once, like GPT-4 and Gemini.  Therefore, we must send each call context method individually and concatenate the summaries into a single list once all methods are processed.

\subsubsection{Target Method Summary Generation with Context}
\label{sec:approach_distil}

This step corresponds to Figure~\ref{fig:overview}, Area 3b.  The \texttt{jam} model provided by Su~\emph{et al.}~\cite{su2024distilled} is able to summarize single Java methods, but it is not able to understand code context since it was only trained to understand the prompt structure in the previous section.  Therefore, we train the model to understand prompts in the style:
\vspace{-1mm}
{
\begin{verbatim}
    TDAT
    {target method}
    CONTEXT
    {descriptions}
    SUMMARY
    {summary}
\end{verbatim}
}
\vspace{-1mm}
We train the model to understand these prompts in a knowledge distillation process similar to how Su~\emph{et al.}~\cite{su2024distilled} trained the code summarization model.  We first collect a dataset of Java methods and the projects from which they originate.  Then we extract the call context for each Java method.  Then we use the procedure we described in Section~\ref{sec:approach_commercial} to obtain a code summary for each Java method using the dataset of methods and call context we extracted.  We created a training set of prompts in the style above in which {\small\texttt{\{target method\}}} is the target method's source code, {\small\texttt{\{descriptions\}}} is the list of summaries we generate for the methods in the target's call context using the procedure in Section~\ref{sec:approach_callcontext}, and {\small\texttt{\{summary\}}} is from the commercial procedure in Section~\ref{sec:approach_commercial}.  Note that we choose which of the commercial models to use (GPT-4 or Gemini) as part of our evaluation in Section~\ref{sec:eval}.

We used a dataset of 170k Java methods recommended by Bansal~\emph{et al.}~\cite{bansal2023function}.  This dataset was already heavily filtered and preprocessed to remove duplicates and other potential problems.  We created a total of 170k training prompts (one for each method) and used these prompts to train the \texttt{jam} model using the recommended default settings for code summarization~\cite{su2023language, su2024distilled} for a total of four epochs.  The result is a model capable of understanding the prompt style above.  Note that this model is a distilled version of one of the commercial models in Section~\ref{sec:approach_commercial} and is not yet ``finished.''  We use this model as a basis for the fine-tuning procedure next.

\subsubsection{Fine-tuning from Curated Examples}
\label{sec:approach_ftcurated}

This step corresponds to Figure~\ref{fig:overview}, Area 3c.  We finetune our model from the previous subsection using a selection of very high-quality examples from a human study on code summarization in context.  However, due to the small number of samples, we implement a custom weight freezing fine-tuning strategy.

\paragraph{Obtaining Exemplary Summaries}

We obtained the examples from Bansal~\emph{et al.}~\cite{bansal2024programmer} who conducted an experiment in which they asked six programmers to write summaries for 40 Java methods each (from five different Java projects).  A novel aspect to their study was that they asked programmers to focus on \emph{why} the methods exist in the program rather than only the internals of each method.  The programmers spent around six hours each on the experiment.  One key outcome of the experiment was a dataset of 236 (40 methods $x$ 6 programmers, with four samples discarded due to study errors) one-sentence summaries of Java methods that focus on why the method exists.  We consider these summaries to be exemplars, since they were created to follow rigorous quality guidelines, after significant manual effort involving expert programmers, in a controlled environment. %YH: the $x$ here needs to be replaced with another format

In that study, Bansal~\emph{et al.} designed an interface in Eclipse, a popular IDE for Java development. They presented the participants with raw source code for one Java project per study session, with comment and other artifacts removed. They presented the participant with paths to 8 methods from various files in the project. They then asked the participants to write exemplar summaries that describe why that method is useful in the context of the project. Each participant was asked to complete five such sessions, each for one of the five projects cleaned for the study. These projects were extracted from GitHub in September 2023, and were created for diverse applications. The participants had access to the source code of the entire project to assess and generate the exemplar summaries.
%YH: I kind of feel this paragraph can be shortened (e.g., do we need to know they designed an interface in Eclipse?) and also you can replace "they did XYZ" with "the dataset was collected via ..." It reads like you spent a lot of space promoting something that is  done and published by others (though those are Aakash's previous work :D).

\paragraph{Weight Freezing}

We froze the weights of some layers of the model to prevent it from overfitting.  To ``freeze weights'' means to prevent some components of the neural model from training during finetuning.  The idea is to preserve part of the model to avoid overfitting to a small number of samples.  While the concept has been proposed several times over the years~\cite{alissa2022performance, kwok1993experimental, goutam2020layerout, malan2023automatic, wimmer2023dimensionality}, there is no single accepted approach; we had to implement it in a novel way.  Our strategy was to freeze the weights of some layers in the model.  Recall that the \texttt{jam} model is based on the GPT-2 architecture, and the model Su~\emph{et al.}~\cite{su2023language} release is a 24-linear-layer and 1,024-embedding-size configuration.  We froze embedding layers and 25\% of the linear layers and only fine-tune the unfrozen layers.  The frozen layers are all embedding layers and the linear layers whose layer number is divisible by four.  Others are trainable during fine-tuning.  Other than weight freezing, we used all other hyperparameters recommended by Su~\emph{et al.}~\cite{su2023language} for code summarization.
%YH: I feel reviewers will ask how would the portion of frozen layers affect the performance. Why do you decide to freeze all embedding layers and 25% of linear layers?

\newpage

\begin{figure}[h!]
    \centering
    \includegraphics[width=0.3\linewidth]{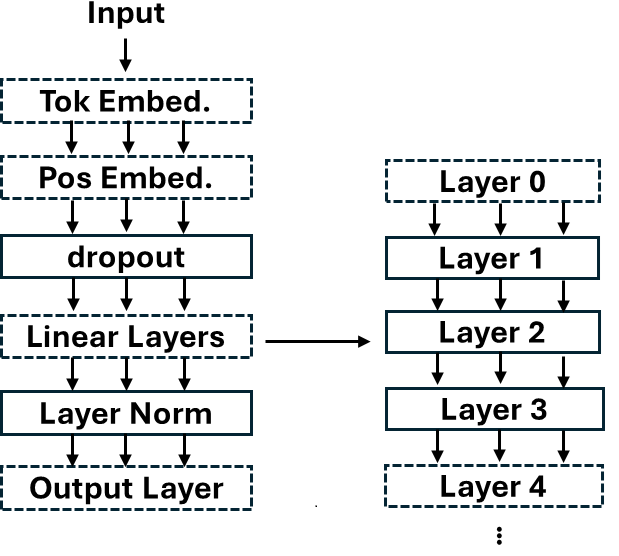}
    \caption{Overview of Weight Freezing. The solid-line and dot-line means the weights are trainable and frozen respectively. }
    \label{fig:freeze_overview}
    \vspace{-4mm}
\end{figure}

As shown in Fig.~\ref{fig:freeze_overview}, we froze the weights of all embedding layers (including token embedding,  position embedding, output embedding) and parts of linear layers. We froze the embedding layers to preserve the embedding information that the model learned during pre-training, because fine-tuning data is relative small compared with the pre-training data. We do not want to confine our embedding to be in the topics of our fine-tuning set only. We froze the output embedding layer due to the weight tying technique~\cite{press2016using} in the model that Su~\emph{et al.}~\cite{su2023language} released. We froze the layers where the layer number is divisible by four because 1) we want the results to be reproducible compared with choosing any layers randomly each time 2) we do not want to solely freeze top or last few layers because Merchant~\emph{et al.} found that fine-tuning can have effects on different layers~\cite{merchant2020what}. Overall, these result in 75m trainable parameters.  Like many approaches using neural models, we decided on this approach after numerous pilot studies and trial-and-error.

%\vspace{-4mm}

%YH: this figure is hard to read - text is too small. And you have a lot of white space in it
%\vspace{-4mm}
Note that in pilot studies we attempted to fine-tune the model using the 256 examples using recommended settings~\cite{su2023language}.  However, since there is significant overlap in the examples (since they cover only 40 methods), we found that the model would often generate outputs too similar to the training data regardless of the actual context at inference time.  We found that the usual techniques to avoid overfitting (e.g., increasing dropout~\cite{srivastava2014dropout}, lowering learning rate~\cite{li2019towards}) did not mitigate the problem.  We chose weight freezing as an alternative, and we chose to freeze every other layer after different configurations (e.g., the first \emph{n} layers, every two layers) yielded inferior results in our subjective opinion.  We do not seek a complete evaluation of avoid overfitting in this paper -- instead, we seek to demonstrate a proof-of-concept approach versus a baseline state-of-the-practice.

\section{Evaluation}
\label{sec:eval}

We evaluate our approach in two conditions.  First, we study our approach using commercial models as described in Section~\ref{sec:approach_commercial}.  Second, we compare our approach with the fine-tuned open-source model (from Section~\ref{sec:approach_opensource}) to the best-performing commercial model and references written by human experts.

\subsection{Research Questions and Setting}

Our research objective is to evaluate our approach using both commercial models and fine-tuned open-source models.  We ask the following Research Questions (RQs) to that end:

\begin{itemize}
\vspace{0.5mm}
\item[RQ1] What is the best commercial model for this task?
\vspace{0.5mm}
%\item[RQ2] How well does the open-source model duplicate the best commercial model after fine-tuning?
\item[RQ2] How do the fine-tuned open-source model and the best commercial model compare to references written by human experts?
\vspace{0.5mm}
\end{itemize}
%YH: I found the phrasing of RQ1 a bit confusing: since you mention configuration, I thought you will talk about things like temperature in GPT models. Even with the explanation below for the 4 setups, the rationale is stil confusing. Cause there seems to be a separate purpose for the 4 setups: area 2 is only for baselines (w/o context), process 3 in the table is one of your two implementations. I am not sure why they are called "configuration" as a whole. it makes me feel that you are doing one type of experiment (e.g., context-aware summarization), but using these 4 setups and want to compare which one is the best and recommend people to use it. Maybe consider the following RQs: (1) How does LLM+context perform on code summarization? here you can introduce baseline setups (2) How feasible is it to do LLM+context code summarization using small LLM models locally? 
% CMC: Fair point, I changed the wording a bit.  I see your overall point but the RQs are closely aligned with the sets of experiments, and RQ2 covers both small LLMs and the reference summaries.

%YH: I have another question here in the evaluation: should you have a baseline that is context + non-LLM models? right now you are only comparing among different LLMs.
% CMC: We found in other papers (cited, but I can do a better job of highlighting them) that LLM approaches completely stomp non-LLM approaches at this time.  It's not even a remotely close comparision (sadly).

The rationale behind RQ1 is that we fine-tune an open-source model to duplicate a commercial model (described in Section~\ref{sec:approach_distil}, prior to fine-tuning the model again using human references, Section~\ref{sec:approach_ftcurated}).  But there are four possible configurations of the commercial models because we implement our approach using one of two models (see Section~\ref{sec:approach_commercial}) and one of two processes:

%Process 1 corresponds to the baseline where the model sees only the source code of the target method itself (shown in Figure~\ref{fig:overview}, Area 1).  
Process 2 corresponds to the baseline where we provide all project source code to the model as one input (Figure~\ref{fig:overview}, Area 2).  Process 3 corresponds to our process for commercial models described in Section~\ref{sec:approach_commercial}.  We do not use Process 1 as a baseline because the model does not see any context, making the task of describing context unrealistic.  In the remainder of this paper, we refer to these configurations using the names shown in the table above.  RQ1 seeks to determine the best of these configurations.

\vspace{-3mm}
\begin{table}[h!]
\centering
\begin{tabular}{lll}
          & GPT-4        & Gemini         \\
Process 2 & \texttt{gpt4-base}    & \texttt{gemini-base}    \\
Process 3 & \texttt{gpt4-context} & \texttt{gemini-context}
\end{tabular}
\end{table}
\vspace{-3mm}

%Process 1 corresponds to the baseline where the model sees only the source code of the target method itself (shown in Figure~\ref{fig:overview}, Area 1).

The rationale behind RQ2 is that we fine-tune the open-source model a second time (Section~\ref{sec:approach_ftcurated}) using reference summaries written by human programmers who were hired to read the whole project's source code and write summaries.  Since these references have highest quality and we fine-tune a model using them, we evaluate how the fine-tuned model compares to these references.  We also compare the best-performing commercial model configuration to the references because it is theoretically possible for the commercial model to outperform human-written references~\cite{su2024distilled}.

\subsection{Hardware / Software Details}

We implemented our approach on a machine with an Intel i9-10900X CPU, four NVidia RTX A5000 GPUs, and 256GB system memory.  Software included Python 3.11, PyTorch 2.2.1, and CUDA 12.2.  The GPT-4 model name we used was {\small\texttt{gpt-4-0125-preview}}.  The Gemini model we used was not named but was current as of March 20, 2024. %YH: NVIDIA RTX A5000? % CMC: Yes.

\subsection{Experiment Method}
\label{sec:eval_method}

Our experimental method focuses on a human study design.  We use a human study in lieu of automated metrics such as BLEU and METEOR because recent language models can generate summaries with a wide variety of wording, and often that wording is equal or even superior to human-written references~\cite{su2024distilled, mastropaolo2024evaluating, ghassemiazghandi2024evaluation, sun2023automatic, yuan2023evaluating}.  The human subjective opinion in these studies has long been considered the gold standard in evaluating code summary generation despite extensive work on metrics~\cite{eddy2013evaluating, haque2022semantic,  mastropaolo2024evaluating, shi2022we, stapleton2020human, roy2021reassessing}.

We use a tournament-style evaluation strategy in which two sources of summaries ``compete'' in each of six experiments with a different set of human participants (groups A-F), as in Table~\ref{tab:tournament}.

\begin{table}[h!]
\caption{Depiction of Tournament Style Evaluation}
\begin{tabular}{llll}
experiment & participants & summary source                                      &                      \\ \cline{1-3}
1    & A                 & \multicolumn{1}{l|}{\small gemini-context vs. gemini-base} & \multicolumn{1}{c}{} \\
2    & B                 & \multicolumn{1}{l|}{\small gpt4-context vs. gpt4-base}     & RQ1                  \\
3    & C                 & \multicolumn{1}{l|}{\small best-of-exp1 vs. best-of-exp2}  &                      \\ \hline
4    & D                 & \multicolumn{1}{l|}{\small jam-ft vs best-of-exp3}       &                      \\
5    & E                 & \multicolumn{1}{l|}{\small references vs. best-of-exp3}  & RQ2                  \\
6    & F                 & \multicolumn{1}{l|}{\small best-of-exp4 vs. best-of-exp5}  &                     
\end{tabular}
\label{tab:tournament}
\end{table}

%YH: can consider indicate the number of participants in each group in the table. Also it is a bit weird this table has no number and caption
% CMC: It's my attempt to keep the page limit under control.  The description is in the surrounding text as the tables are supportive of discussion in that text ("following table:").  We've done this before and no review has complained about it. :-)  The text does say that each group has ten participants.

\begin{figure}[b!]
    \centering
    \includegraphics[width=1.0\linewidth]{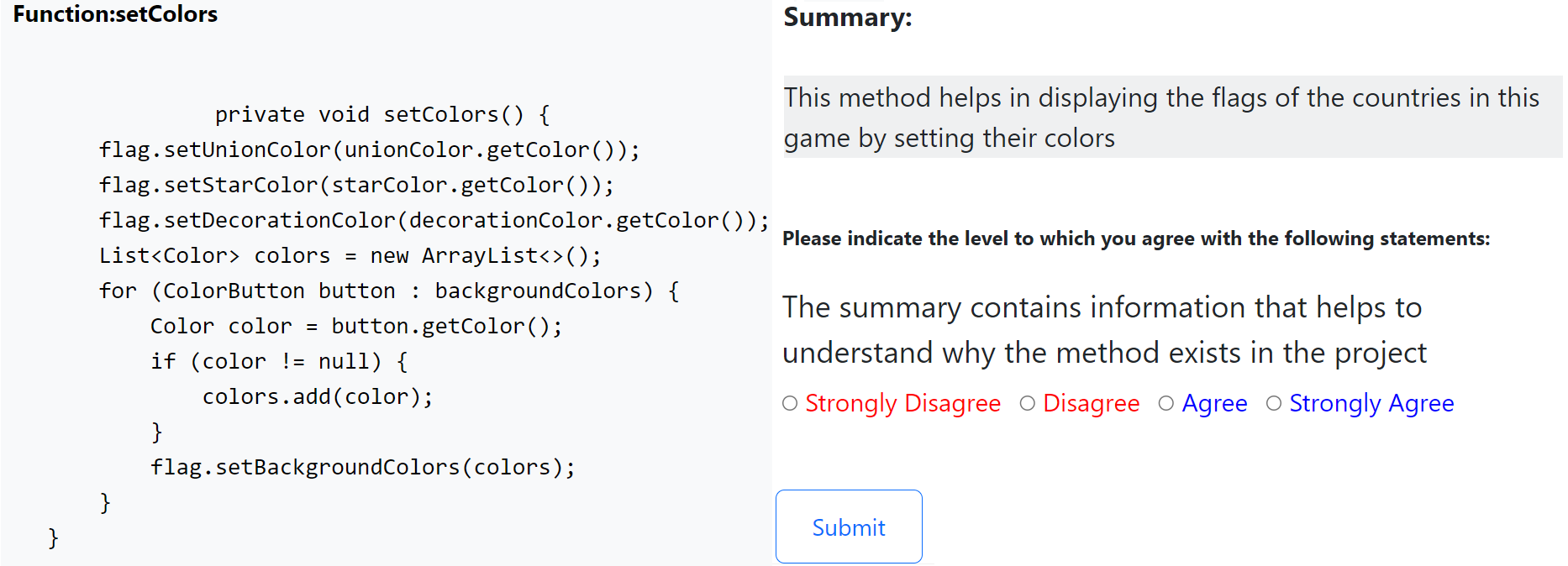}
    \includegraphics[width=1.0\linewidth]{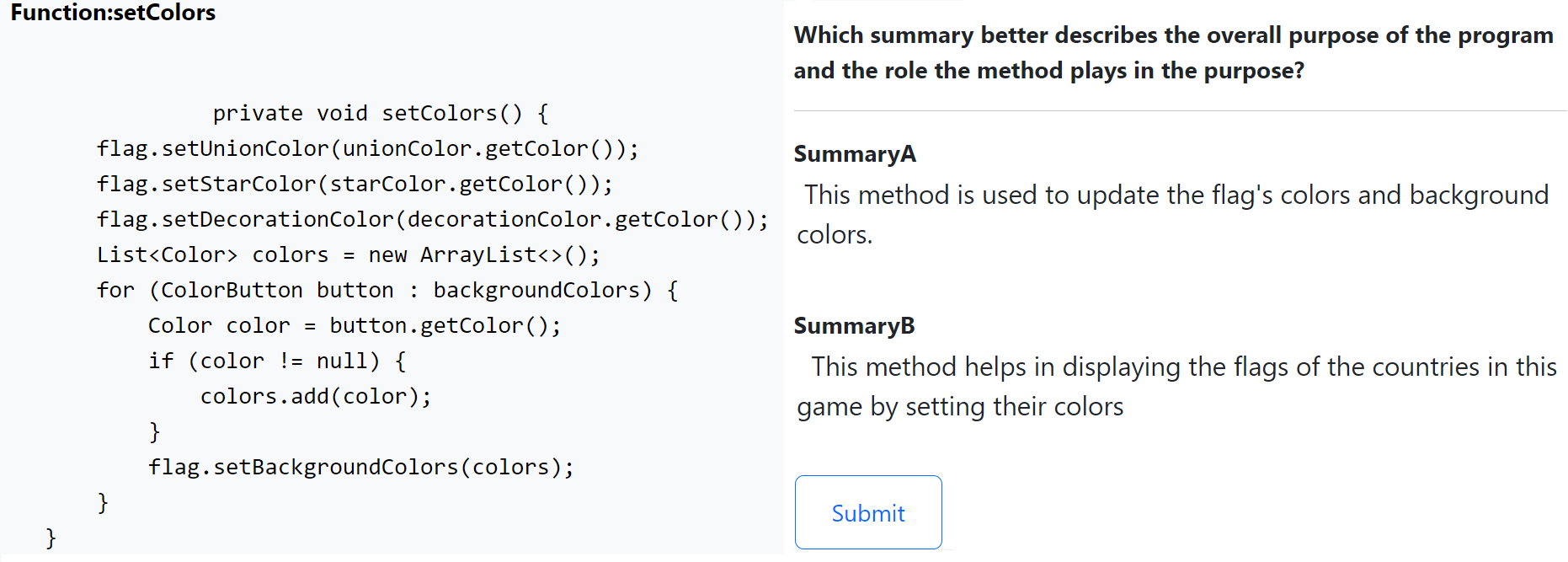}
    \caption{The interface for our study.  Each participant saw two pages for each Java method.  The top image shows the first page and the bottom image shows the second page.}
    \label{fig:survey}
\end{figure}
%YH: Figure 3 is not readable yet. 

For RQ1, we compare {\small\texttt{gemini-context}}, {\small\texttt{gemini-base}}, {\small\texttt{gpt4-context}}, and {\small\texttt{gpt4-base}}.  In experiments 1 and 2 we compare the base to the context versions using each model, then in experiment 3 we compare the ``winner.''

The experiments for RQ2 depend on the outcome of the RQ1 experiments: we fine-tune the open-source model to be a distilled version of the best of experiment 3 (recall Section~\ref{sec:approach_distil}) then further fine tune it using human-written exemplar summaries (recall Section~\ref{sec:approach_ftcurated}).  We refer to the finished open-source model as \textbf{\texttt{jam-ft}}.  Then, in experiment 4, we compare this model to the model from which it was distilled.  We also, in experiment 5, compare that model to the exemplar human-written summaries using a leave-one-out validation strategy, which we will explain in Section~\ref{sec:eval_leaveoneout}.  %Finally, in experiment 6, we compare the best of experiments 4 and 5.

Each experiment consists of a web-based study.  We designed our study based on previous studies by McBurney~\emph{et al.}~\cite{mcburney2014automatic, mcburney2016automatic}, Haque~\emph{et al.}~\cite{haque2022semantic}, and Sridhara~\emph{et al.}~\cite{sridhara2011automatically}.  We use a web interface to show the source code and summaries for a series of Java methods to programmers, who then answer questions about the summaries.  Figure~\ref{fig:survey} shows the interface.  We show two pages for each Java method.  One page shows the code of the method and a summary, followed by the statement:

%YH: please consider highlighting the purpose of each question as /paragraph{} or bold text -  What are  you checking behind each question?

\begin{itemize}
    \item[] Question 1: ``The summary contains information that helps to understand why the method exists in the project.''
\end{itemize}

Then options for ``Strongly Disagree'', ``Disagree'', ``Agree'', ``Strongly Agree'' corresponding to a four-point Likert scale recommended in related work~\cite{mcburney2014automatic, sridhara2011automatically}.  This question results in a 1-4 rating for each summary, the mean of which we compare using a Mann-Whitney statistical test in the procedure recommended by McBurney~\emph{et al.}~\cite{mcburney2014automatic} and present in Table~\ref{tab:stats} in Section~\ref{sec:eval}.  After participants answered this question and clicked submit, the survey showed a second page with the Java method code plus the question:

\begin{itemize}
    \item[] Question 2: ``Which summary better describes the overall purpose of the program and the role the method plays in the purpose?''
\end{itemize}

Followed by two summaries.  The participants select one of the summaries by clicking on it and the clicking the submit button.  One of the two summaries is the one from the first page.  The second is the ``competitor'' summary.  Note that in any given study we are comparing two summary sources.  The survey randomly selects a summary as the first one to show (on page one) and then the other alongside it on page two.  We derived this question from an experiment by Su~\emph{et al.}~\cite{su2024distilled} that compared code summaries from commercial language models to distilled open-source models.
%YH: I don't think it is clear what is the competitor summary and where it s from. Maybe refer back to the table at the beginngin of the C subsection?

\subsection{Avoiding Common Biases}
\label{sec:eval_biases}
%YH: if you need space, can shorten the bias part and integrate the quality control part together.
We designed our study to avoid the following common biases with human studies in software engineering:

\emph{Fatigue Bias.}  Fatigue bias occurs in human studies when the participant has worked at the task for an extended period of time and may give lower quality or higher/lower ratings than when not fatigued~\cite{ko2015practical}.  We avoid fatigue bias by limiting each experiment to 40 Java methods (80 total pages) which allows participants to complete each experiment in about one hour.

\emph{Demand Characteristic Bias.}  Dell~\emph{et al.}~\cite{dell2012yours} point out that a major bias in software engineering studies occurs when participants are aware of the experimenter's stake in the study.  We do not display which summary is from which approach, and we select the order of the summaries randomly.

\emph{Leading Bias.}  This bias occurs when participants are pressured to answer positively or negatively to a question~\cite{romano2021researcher}.  To avoid it, we used a statement from experiments in related work instead of a question on the first page.  And on the second page, we asked participants to select a summary in response to a question instead of selecting a level of agreement.

\emph{Instrument Bias.}  This bias is from the study interface itself.  A key instrument bias when we show source code is that coloration and formatting of source code.  We used a plain, non-syntax-highlighted font when displaying code to avoid this bias, as recommended by McBurney and McMillan in their previous study~\cite{mcburney2014automatic}.

\subsection{Participants}

We recruited 60 participants for our study.  We divided the participants into six groups of ten people each (groups A-F, one group for each experiment).  We recruited using the Prolific platform~\cite{prolific} with the criteria that participants must be 25 years of age, be located and authorized to work in the United States or United Kingdom (for legal and English proficiency reasons), hold a degree in Computer Science, and have at least one year of Java experience.  We offered \$15/hr in compensation, which is a strong rate on the platform. %We removed four participants' results for quality control using the procedures in the next section but ensured a minimum of ten per experiment.

\newpage

\subsection{Quality Control}

Two problems often occur in web-based studies and these problems necessitate quality control.  First, results may be inconsistent because of the impossibility to have an introduction presentation to all participants as in the in-person studies.  Second, persons with ill motives or without programming experience may enter fraudulent results to receive survey compensation without significant effort.  The typical mitigation strategy is to use screening tests to ensure programming proficiency~\cite{danilova2021you}, but recent work pointed out that the rise of AI-powered tools has made these tests ineffective~\cite{ghorbani2023autonomy}.

Instead, we built seven sample questions into the study that resemble real questions, a strategy we adapted from related work~\cite{ghorbani2023autonomy}.  We designed these questions to have clear answers in line with our expectations.  We showed four of these samples to participants at the beginning of the study as part of instructions on how to do the study.  Then we randomly included three samples in the study as quality control questions.  The participant had to get all three correct in order for the results to be valid.  We informed the participants that these three would be in the study, but did not inform them which questions were for quality control.  In the end, we removed four participants' results for quality control using the procedures in this section but ensured a minimum of ten per experiment (i.e., we recruited a total of 64 participants but only 60 were valid).

\subsection{Subject Java Methods}

We used the set of 40 Java methods from Section~\ref{sec:approach_ftcurated} in all of our experiments, using a leave-one-out procedure described in the next section to avoid data contamination.  We used these 40 methods because: 1) they have been curated to ensure very high quality of the samples, 2) we have confidence that the summaries are not in the training sets of commercial language models because they have not been uploaded to public repositories yet, 3) they represent a diverse set of applications, and 4) more than 40 samples leads to fatigue bias (see Section~\ref{sec:eval_biases}).  We used all 40 in all six experiments to help ensure consistency of the results, but randomized the order in which the participants saw the methods.  The statistics for the method and the summary size (as tokenized by GPT-2) are shown in Table~\ref{tab:tokstatistics}.

\vspace{-1mm}

\begin{table}[h!]
\centering
\caption{Statistics for method and summary }
\begin{tabular}{p{55mm}l}
mean number of tokens per method         & 291 \\
maximum number of tokens per method      & 958 \\
minimum number of tokens per method      & 87  \\
mean number of tokens in summaries       & 20  \\
mean number of functions in call context & 3  
\end{tabular}
\label{tab:tokstatistics}
\end{table}
%YH: It it a bit rare to see a table without a caption and hlines

%Now we need some statistics on the length of the methods and projects and other information.  Lorem ipsum dolor sit amet, consectetur adipiscing elit, sed do eiusmod tempor incididunt ut labore et dolore magna aliqua. Ut enim ad minim veniam, quis nostrud exercitation ullamco laboris nisi ut aliquip ex ea commodo consequat. Duis aute irure dolor in reprehenderit in voluptate velit esse cillum dolore.

% number of projects: 5
% number of tokens in average: 291
% maximum number of tokens: 958
% minimum number of tokens: 87
% average number of tokens in summaries: 20
% average number of the functions that calls the function: 3

\subsection{Leave-one-out Reference Comparison}
\label{sec:eval_leaveoneout}

We use a leave-one-out evaluation procedure for the {\small\texttt{jam-ft}} approach in Experiment 6 when comparing {\small\texttt{jam-ft}} to the references.  What this means is that when comparing {\small\texttt{jam-ft}} to a reference summary for a Java method, we train {\small\texttt{jam-ft}} only with the other 39 methods so that the model does not see the example under evaluation.  This procedure is necessary because the number of exemplar samples we have is small due to the exorbitant expense in creating these samples.  It is not feasible to train the model with dozens of samples and then also set aside dozens for evaluation only.  This situation is common in software engineering experiments because curated source code samples have historically been limited, and various studies recommend using a leave-one-out strategy as a solution~\cite{ali2021empirical, cleland2007automated, kocaguneli2013software, tantithamthavorn2016empirical}.  Note that the {\small\texttt{jam-ft}} model in Experiment 4 is trained with all 40 exemplar samples because there are no references used in Experiment 4 (which is versus {\small\texttt{gemini-context}} not references).  We release all models in our reproducibility package (see Section~\ref{sec:reproducibility}), but recommend the use of the model from Experiment 4 because it includes the maximum number of training samples.

%YH: though it is OK, it is more commonly seen to put threat to validity before conclusion
% CMC: Really?  I've always put them at the end of the evaluation methods discussion.  I may be weird.

\vspace{-2mm}

\section{Evaluation Results}

This section seeks to answer our research questions through analysis of the results from the experiments in the previous section.

\subsection{RQ1: Commercial Model Performance}

\begin{figure}[h]
    \centering
    \begin{tabular}{ll}
    \includegraphics[width=0.37\linewidth]{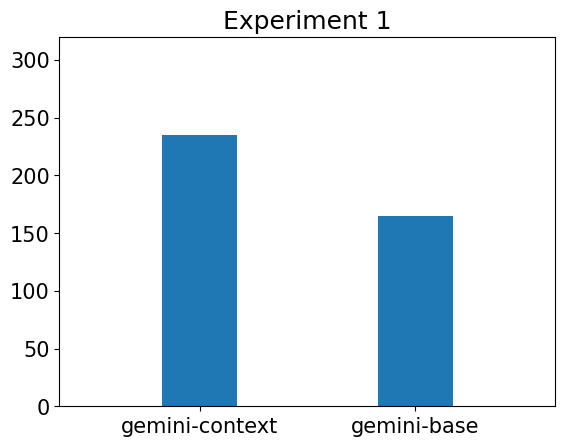} & 
    \includegraphics[width=0.37\linewidth]{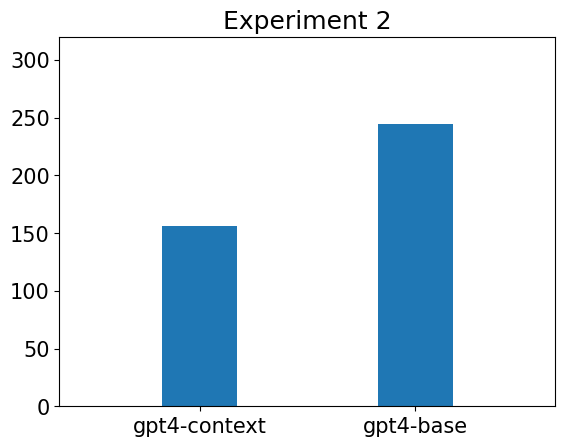} \\
    \includegraphics[width=0.37\linewidth]{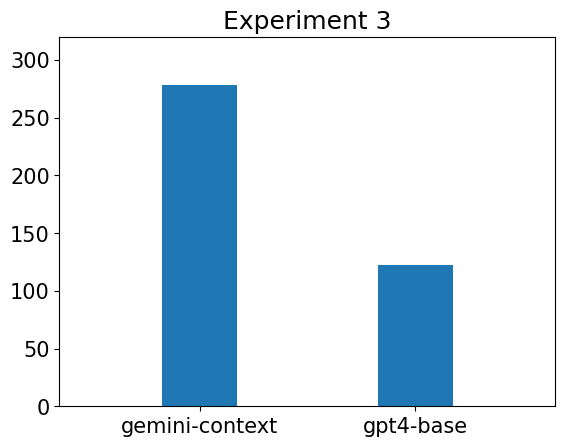} & 
    \includegraphics[width=0.37\linewidth]{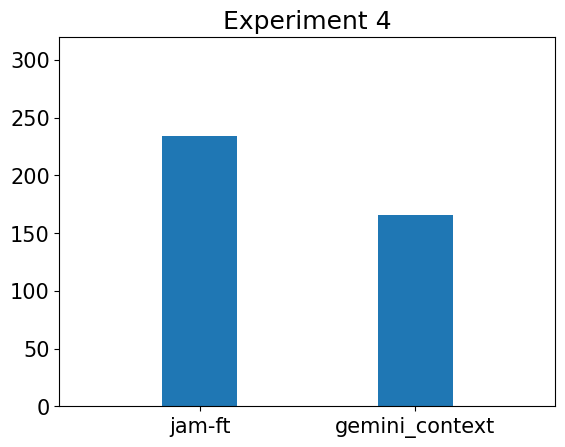} \\
    \includegraphics[width=0.37\linewidth]{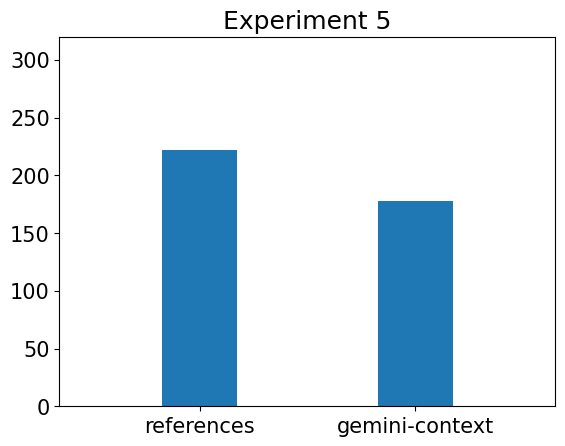} & 
    \includegraphics[width=0.37\linewidth]{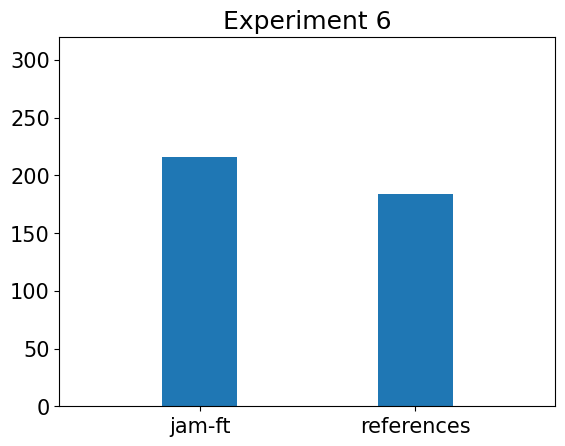}
    \end{tabular}
    \caption{Bar charts showing the responses to Question 2 for each experiment.  For example, in Experiment 3, there were approximately 275 samples for which participants preferred {\small\texttt{gemini-context}} versus about 125 for {\small\texttt{gpt4-base}}.}
    \label{fig:exp_q2_results}
  
\end{figure}

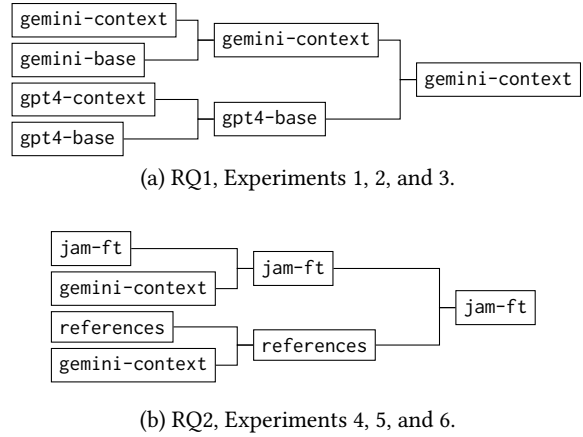
\begin{figure}[h!]

\centering
\begin{tikzpicture}%
[ every text node part/.style={draw, align = left, inner sep = 0pt} ]

% Setup for horizontal tree
    % Grow tree to right with nodes placed clockwise(')
    \tikzset{grow'=left}
    % Use edges with 90° bends instead of default straight
    \tikzset{edge from parent/.style = { draw,
             edge from parent path = { (\tikzparentnode.west) 
                                        -- +(-6pt, 0)
                                        |- (\tikzchildnode.east) }}}
    % Increase horizontal spacing (adjust if length of name is long)
    \tikzset{level distance = 8.5em}
    % Adjust the alignment of the nodes
    \tikzset{every tree node/.style = {draw, anchor = base west}}

\Tree[ .{{\small\texttt{gemini-context}}}
        [ .{{\small\texttt{gpt4-base}}} {{\small\texttt{gpt4-base}}} {{\small\texttt{gpt4-context}}} ]
        [ .{{\small\texttt{gemini-context}}} {{\small\texttt{gemini-base}}} {{\small\texttt{gemini-context}}} ] ]
\end{tikzpicture}

(a) RQ1, Experiments 1, 2, and 3.

\vspace{5mm}

\begin{tikzpicture}%
[ every text node part/.style={draw, align = left, inner sep = 0pt} ]

% Setup for horizontal tree
    % Grow tree to right with nodes placed clockwise(')
    \tikzset{grow'=left}
    % Use edges with 90° bends instead of default straight
    \tikzset{edge from parent/.style = { draw,
             edge from parent path = { (\tikzparentnode.west) 
                                        -- +(-6pt, 0)
                                        |- (\tikzchildnode.east) }}}
    % Increase horizontal spacing (adjust if length of name is long)
    \tikzset{level distance = 8.5em}
    % Adjust the alignment of the nodes
    \tikzset{every tree node/.style = {draw, anchor = base west}}

\Tree[ .{{\small\texttt{jam-ft}}}
        [ .{{\small\texttt{references}}} {{\small\texttt{gemini-context}}} {{\small\texttt{references}}} ]
        [ .{{\small\texttt{jam-ft}}} {{\small\texttt{gemini-context}}} {{\small\texttt{jam-ft}}} ] ]
\end{tikzpicture}
\vspace{2mm}

(b) RQ2, Experiments 4, 5, and 6.

\caption{Tournament bracket showing experiment outcomes for each research question.  The ``winners'' in these brackets are determined based on which model received more votes for Question 2 in the survey (see Section~\ref{sec:eval_method}).}
\label{fig:bracket}
\vspace{4mm}
\end{figure}

We find that the {\small\texttt{gemini-context}} model was the best performer of the commercial models.  Our evidence to support this finding comes from Experiments 1 through 3.  Consider the tournament bracket in Figure~\ref{fig:bracket}(a).  We compared {\small\texttt{gemini-context}} to {\small\texttt{gemini-base}} in Experiment 1, and {\small\texttt{gemini-context}} had more times that participants preferred summaries from that model than from the base version, as measured by responses to Question 2 during that experiment.  The difference is visible in Figure~\ref{fig:exp_q2_results}.  In contrast, for Experiment 2, {\small\texttt{gpt4-base}} outperformed {\small\texttt{gpt4-context}} in terms of responses to Question 2.  Each experiment had a total of 400 responses to Question 2 (40 methods times 10 participants), and {\small\texttt{gpt4-base}} was preferred around 250 times compared to 150 for {\small\texttt{gpt4-context}}.

Note that because both GPT4 and Gemini are closed-source, it is not possible to firmly state why one model outperforms another.  Nonetheless, we observe that Gemini seems more able to process code context when generating summaries in our experimental settings.  In addition, in Experiment 3, we observed that {\small\texttt{gemini-context}} was preferred to {\small\texttt{gpt4-base}}.  We discuss representative examples and potential reasons behind these findings in Section~\ref{sec:examples}.

%\newlength{\originaltabcolsep}
%\setlength{\originaltabcolsep}{\tabcolsep}
%\newcommand{\originalarraystretch}{\arraystretch}

%\setlength{\tabcolsep}{1pt}
%\renewcommand{\arraystretch}{2.0}

Additional evidence is from the responses to Question 1.  While Question 2 is a direct comparison of the summaries in terms of the purpose of the program and role the method plays, Question 1 asks for agreement to a statement about the summary's content about why the method exists (recall Section~\ref{sec:eval_method} for precise wording).  The responses of Question 1 are in the form of a four-point Likert scale, for which we present a comparison of means for each study in Table~\ref{tab:stats}.  In Experiment 1, {\small\texttt{gemini-context}} received a higher mean score by a statistically significant margin compared to {\small\texttt{gemini-base}}.  Likewise, in Experiment 2, {\small\texttt{gpt4-base}} outperformed {\small\texttt{gpt4-context}}, and {\small\texttt{gemini-context}} outperformed {\small\texttt{gpt4-base}} in Experiment 3, both by statistically significant margins.

Roy~\emph{et al.}~\cite{roy2021reassessing} point out that statistically significant differences from Likert scale survey responses for source code summarization are a high bar to cross due to variations in human subjective ratings.  According to Roy~\emph{et al.}, statistical significance is often only observed when the differences between summaries are large, and smaller differences may present but not detectable by coarse Likert scale ratings.  The direct comparison of summaries is suggested by Su~\emph{et al.}~\cite{su2024distilled} as a more sensitive alternative, which we use in Question 2.  Therefore, in answering RQ1, we consider the findings of Question 1 to be a strong support of the conclusion that {\small\texttt{gemini-context}} is the best performing baseline using a commercial model.

\subsection{RQ2: Open-source Model and References}

We find that the {\small\texttt{jam-ft}} model produced the highest performing summaries overall.  Recall from Section~\ref{sec:approach_distil} that we create {\small\texttt{jam-ft}} by using the distillation process described by Su~\emph{et al.}~\cite{su2024distilled} from the summaries generated by the best-performing commercial model.  We determined that the best commercial model to be {\small\texttt{gemini-context}} and therefore used it in the procedures described in Sections~\ref{sec:approach_distil}~and~\ref{sec:approach_ftcurated} to create the {\small\texttt{jam-ft}} for RQ2.

Consider Figures~\ref{fig:exp_q2_results}~and~\ref{fig:bracket}(b), which shows the outcome of Experiments~4, 5, and 6.  In Experiment~5, we observe that the quality of human references was perceived to be higher than the summaries from {\small\texttt{gemini-context}}.  This result is not surprising, because the references were written by people who were specifically hired to write high quality summaries after reading the source code and context.  However, we note that the margin is not as high as we observe for experiments for RQ1: the difference visible in Figure~\ref{fig:exp_q2_results} for Question 2 is not as large and the difference in Question 1 is not statistically significant. (see Table~\ref{tab:stats}).  Likewise, in Experiment~4, {\small\texttt{jam-ft}} outperforms {\small\texttt{gemini-context}}, but also by a margin smaller than that observed in RQ1 experiments.  This result may also be expected given that {\small\texttt{jam-ft}} was trained to mimic {\small\texttt{gemini-context}} and then further trained with human-generated examples.

But to our surprise, {\small\texttt{jam-ft}} slightly outperformed the references written by human experts, according to Question~2 responses in Experiment~6.  Approximately 215 summaries from {\small\texttt{jam-ft}} were rated as higher quality than the references, versus 185 references rated as higher than {\small\texttt{jam-ft}}.  However, we note that differences in Question~1 responses were not statistically significant.  Therefore, we cannot strongly claim that {\small\texttt{jam-ft}} produces better summaries than human experts, though we do observe evidence that the quality is at least perceived as similar, which is a good result considering the low cost of the model compared to human involvement.

In the end, we conclude that {\small\texttt{jam-ft}} produced summaries rated as the best in our experiments.  We explore potential reasons for this finding with examples in the next section.

\begin{table}[!h]
\caption{Mann-Whitney statistical test results for page~1 for each experiment (1-3 for RQ1 and 4-6 for RQ2).  Note: means are only comparable within experiments, not across experiments, because each experiment had different participants.}
\vspace{1mm}
\small
\label{tab:stats}

\begin{tabular}
{llp{4mm}p{4mm}p{4mm}p{4mm}p{4mm}}
\small Experiment                              &  Source         & Mean  & Stddev & Zobs                   &  Zcrit                  & p                                \\ \hline
\multicolumn{1}{c|}{\multirow{2}{*}{1}} &  gemini-base    & 2.511 & 0.843  & \multirow{2}{*}{1.679} & \multirow{2}{*}{1.645} & \multirow{2}{*}{0.047}           \\
\multicolumn{1}{c|}{}                   & gemini-context & 2.636 & 0.867  &                        &                        &                                  \\
\multicolumn{1}{c|}{\multirow{2}{*}{2}} & gpt4-base      & 2.506 & 0.780  & \multirow{2}{*}{2.256} & \multirow{2}{*}{1.645} & \multirow{2}{*}{0.012}           \\
\multicolumn{1}{c|}{}                   & gpt4-context   & 2.332 & 0.872  &                        &                        &                                  \\
\multicolumn{1}{c|}{\multirow{2}{*}{3}} & gemini-context & 2.871 & 0.719  & \multirow{2}{*}{3.662} & \multirow{2}{*}{1.645} & \multirow{2}{*}{\textless{}0.01} \\
\multicolumn{1}{c|}{}                   & gpt4-base      & 2.600 & 0.775  &                        &                        &                                  \\ \hline
\multicolumn{1}{c|}{\multirow{2}{*}{4}} & jam-ft         & 2.712 & 0.744  & \multirow{2}{*}{1.180} & \multirow{2}{*}{1.645} & \multirow{2}{*}{0.119}           \\
\multicolumn{1}{c|}{}                   & gemini-context & 2.620 & 0.728  &                        &                        &                                  \\
\multicolumn{1}{c|}{\multirow{2}{*}{5}} & gemini-context & 2.533 & 0.775  & \multirow{2}{*}{1.369} & \multirow{2}{*}{1.645} & \multirow{2}{*}{0.085}           \\
\multicolumn{1}{c|}{}                   & references     & 2.622 & 0.742  &                        &                        &                                  \\
\multicolumn{1}{c|}{\multirow{2}{*}{6}} & jam-ft         & 2.685 & 0.693  & \multirow{2}{*}{0.007} & \multirow{2}{*}{1.960} & \multirow{2}{*}{0.994}           \\
\multicolumn{1}{c|}{}                   & references     & 2.685 & 0.645  &                        &                        &                                 
\end{tabular}
\end{table}

\begin{table}[h!]
%\centering
\vspace{-3mm}
\begin{tabular}{ll}
\textbf{Example 1} & Method ID 19 \\\hline
\end{tabular}

\begin{tabular}{c}
\begin{lstlisting}[language=Java, basicstyle=\ttfamily\footnotesize, breaklines=true, showstringspaces=false, keywordstyle=\color{black}]
private void updateUnitPath() {
 final Unit active = getActiveUnit();
 if (active == null) return;

 Location destination = active.getDestination();
 PathNode path = null;
 if (destination != null
  && !((FreeColGameObject)destination).isDisposed()
  && !active.isAtLocation(destination)) {
  try {
    path = active.findPath(destination);
  } catch (Exception e) {
    logger.log(Level.WARNING, ""Path fail"", e);
    active.setDestination(null);
  }
 }
 setUnitPath(path);
}
\end{lstlisting} \\
\end{tabular}

\begin{tabular}{l|p{6cm}}
\hline
     gemini-context & This method is used to update the unit path after a goto action is stopped, started, or traversed. \\ \cline{2-2}
     jam-ft & This method helps to update the unit's path when the player is moving to a new location.
\end{tabular}
%\caption{Code in Table Cell}
\end{table}
\begin{table}[t]
\vspace{-3mm}

\begin{tabular}{ll}
\textbf{Example 2} & Method ID 6 \\\hline
\end{tabular}

\begin{tabular}{c}
\begin{lstlisting}[language=Java, basicstyle=\ttfamily\footnotesize, breaklines=true, showstringspaces=false, keywordstyle=\color{black}]
protected boolean getStats(Instances instances, double[] stats) {
 stats[0] = stats[1] = stats[2] = 0;
 if (instances.size() == 0) {
   return true;
 }
 double firstTarget = instances.get(0).getTarget();
 boolean stdIs0 = true;
 for (Instance instance : instances) {
   double weight = instance.getWeight();
   double target = instance.getTarget();
   stats[0] += weight;
   stats[1] += weight * target;
   if (stdIs0 && target != firstTarget) {
     stdIs0 = false;
   }
 }
 stats[2] = stats[1] / stats[0];
 if (Double.isNaN(stats[2])) {
   stats[2] = 0;
 }
 return stdIs0;
}
\end{lstlisting} \\
\end{tabular}

\begin{tabular}{l|p{6cm}}
\hline
     gemini-context & This method is used to calculate statistical values (sum of weights, sum of weighted targets, and mean) for a given dataset, considering weight and target values while identifying whether the dataset has a single unique target value. \\ \cline{2-2}
     jam-ft & This method is used when creating a tree with a specified number of instances.
\end{tabular}
\vspace{-4mm}
\end{table}

\subsection{Examples}
\label{sec:examples}

This section presents three examples for comparing {\small\texttt{jam-ft}} and {\small\texttt{gemini-context}} as a basis for potential explanations for the outcome of RQ2.  Examples~1~and~2 show cases when {\small\texttt{jam-ft}} was rated as higher than {\small\texttt{gemini-context}} for both Question~1 and Question~2 in Experiment~4.  Example~3 shows a case when {\small\texttt{gemini-context}} outperformed {\small\texttt{jam-ft}}.

Consider first Example~1.  The context of this Java method is a game program that involves player units on a virtual board.  Observe that the summary from {\small\texttt{jam-ft}} mentions high-level rationale information about the player moving to a new location, versus the summary from {\small\texttt{gemini-context}} which uses project-specific jargon such as ``goto action.''  This is an example of how both summaries can be correct, but the higher-level information is perceived as more useful in understanding the purpose of the method.

\begin{table}[t]
\vspace{-1mm}
\begin{tabular}{ll}
\textbf{Example 3} & Method ID 16 \\ \hline
\end{tabular}

\begin{tabular}{c}
\begin{lstlisting}[language=Java, basicstyle=\ttfamily\footnotesize, breaklines=true, showstringspaces=false, keywordstyle=\color{black}]
private void setColors() {
  flag.setUnionColor(unionColor.getColor());
  flag.setStarColor(starColor.getColor());
  flag.setDecorationColor(decorationColor.getColor());
  List<Color> colors = new ArrayList<>();
  for (ColorButton button : backgroundColors) {
    Color color = button.getColor();
      if (color != null) {
        colors.add(color);
      }
      flag.setBackgroundColors(colors);
  }
}
\end{lstlisting} \\
\end{tabular}

\begin{tabular}{l|p{6cm}}
\hline
     gemini-context & This method is used to update the flag's colors and background colors. \\ \cline{2-2}
     jam-ft & This method is useful when changing the color of a button or a background color.
\end{tabular}
\end{table}

Example~2 demonstrates a similar tendency in both models.  The {\small\texttt{gemini-context}} summary is technically accurate, but it only focuses on project-specific details, such as lists of types of statistical weights.  The summary from {\small\texttt{jam-ft}}  mentions the high level rationale about when the method is used: when creating a tree.  The context of the method includes several methods for creating and modifying trees. Those are all from a program for searching and sorting database entries.  Studies have for years pointed out that the benefit of this high-level information is in helping the reader quickly grasp what the method does as part of a whole program, without the need to read too many project-specific details~\cite{roehm2012professional, stapleton2020human, maalej2014comprehension, wang2022documentation}.  The human references we use to fine-tune {\small\texttt{jam-ft}} help the model generate summaries with this high-level information.

Example~3 demonstrates a similar pattern, but in a case where {\small\texttt{gemini-context}} was rated better.  As with the other examples, both summaries are technically accurate: the context is a GUI program with various common GUI elements, including notification flags.  The notification flags are important parts of the program, and the placement of a flag can change the way other GUI elements look.  The {\small\texttt{jam-ft}} summary mentions details of changing the button or background color, but the {\small\texttt{gemini-context}} summary points to higher-level information about changing that flag itself.

The same pattern may hint at why {\small\texttt{gemini-context}} outperformed {\small\texttt{gpt4-base}}.  While GPT-4 tended to be consistent in format and phrase summaries in a confident manner without qualifications, it also tended to focus on many low-level details, and the human evaluators detected this difference in the experiments.  The result is that GPT-4 summaries may be correct and in a good format, but not include as much high-level rationale information.  A similar phenomenon has been reported in studies of ChatGPT in other domains~\cite{gilardi2023chatgpt, kocon2023chatgpt}.

\subsection{Threats to Validity}

Like all studies, our experiments carry threats to validity.  Key threats to validity include several that affect similar studies, such as the subject source code, the selection of participants, and the effects of the interface itself.  We attempted to mitigate these threats by using a dataset carefully curated by related work that represents a range of program types, recruiting a relatively large number of participants (60), and efforts to reduce common biases (see Section~\ref{sec:eval_biases}).  However, the threat remains that changes in any of these factors could still affect the outcome of the study.  These factors can be both internal threats in that the results of the study may change, and external threats in that the use of e.g. Java methods may reduce generalizability to other languages.

Two threats to validity unique to our evaluation design include: 1) our leave-one-out procedure, and 2) it is not feasible for programmers to read the entire context when evaluating code summaries.  The leave-one-out procedure allows us to evaluate against all samples in a limited dataset, but it does carry the risk that a slightly different model is used for each sample because each is trained on 39 methods instead of 40.  Also note that we only asked participants in our study questions related to their perception of the usefulness of the summaries -- we did not ask them to read the entire code context due to time constraints (such a study would take many hours per participant).  This reflects a realistic scenario in that people using these summaries would also not read the entire context (the purpose of the summary is to avoid the need to read everything themselves).  However, it means that we cannot ask as many questions as possible to evaluate the quality of every detail in the summaries.  Instead, we rely heavily on the study from which we obtained the samples (see Section~\ref{sec:approach_ftcurated}) to ensure high quality of references, and in our study compare participants' perceptions of model output against those references.

\section{Conclusion}
\label{sec:reproducibility}

This paper advances the state of the art with a novel source code summarization process.  This process is novel for two key reasons.  First, we include code context as part of the input to the model.  We include that context in the form of natural language summaries of the functions or methods that call the target method for which we write a summary.  The reason we use summaries of context is so that we can keep the size of the prompt relatively short.  The advantage of keeping the prompt short is that we can use language models small enough to fit on a single GPU to avoid cost and privacy concerns with large commercial models.  However, our process is capable of using commercial models if desired.

Second, as part of our process for open-source models, we fine-tune the model using high-quality summaries written by human programmers that include information about the code context.  These human-written summaries are very specialized and required many hours to create~\cite{bansal2024programmer}, so we are limited in the number we can obtain.  However, we demonstrated how we can improve the model through fine-tuning even with this small number of samples.  We used a knowledge distillation process to create a strong base to fine-tune, and then used a weight freezing approach to avoid overfitting on the small number of samples.

%We conducted an evaluation in which we compared: 1) different commercial baselines, and 2) the best commercial baseline, human-written references, and our fine-tuned open-source model.  Sixty human programmers participated in our experiments.  We showed how our approach generated summaries that included more rationale information about the Java methods.  

A key implication of our work is that language models can be trained to describe more high-level ``why'' information about source code, when provided some details about other parts of the program in which the code resides.  As an academic laboratory with resources not comparable to industry, we faced several limitations: we had only a few dozen exemplar samples, we used only methods connected via call relationships, and we used a relatively small language model (350m parameters).  Yet, our approach outperformed large commercial models on this specific task.  Our work may be a guide for designers of these larger models to help those models produce code summaries with more information about code context.

We encourage reproducibility and release all code artifacts, datasets, and experimental materials via an online appendix:

\vspace{1mm}
https://github.com/apcl-research/jam-contextsum

%{\small \emph{(Above link anonymized for review but will be released openly.)}}

%\begin{enumerate}
%    \item A
%\end{enumerate}

\section*{Acknowledgements}
This work is supported in part by NSF CCF-2100035 and CCF-2211428. Any opinions, findings, and conclusions expressed herein are the authors and do not necessarily reflect those of the sponsors.

\bibliographystyle{ACM-Reference-Format}
\bibliography{biblio}

\end{document}